\documentclass[conference]{IEEEtran}
\IEEEoverridecommandlockouts
\usepackage[pdftex]{graphicx}
\usepackage{cite}
\usepackage{amsmath,amssymb,amsfonts}
\usepackage{algorithmic}
\usepackage{textcomp}
\usepackage{xcolor}
\usepackage{bm}
\def\BibTeX{{\rm B\kern-.05em{\sc i\kern-.025em b}\kern-.08em
    T\kern-.1667em\lower.7ex\hbox{E}\kern-.125emX}}
\newcommand{\ctext}[1]{\raise0.2ex\hbox{\textcircled{\scriptsize{#1}}}}
\begin{document}

\title{\huge DDSupport: Language Learning Support System that Displays Differences and Distances from Model Speech
}

\author{\IEEEauthorblockN{Kazuki Kawamura\IEEEauthorrefmark{1},
Jun Rekimoto\IEEEauthorrefmark{2}}
\IEEEauthorblockA{The University of Tokyo, Tokyo, Japan\\
Sony CSL Kyoto, Kyoto, Japan
\\
Email: \IEEEauthorrefmark{1}kwmr@acm.org,
\IEEEauthorrefmark{2}rekimoto@acm.org}}

\maketitle

\begin{abstract}
When beginners learn to speak a non-native language, it is difficult for them to judge for themselves whether they are speaking well.
Therefore, computer-assisted pronunciation training systems are used to detect learner mispronunciations. 
These systems typically compare the user's speech with that of a specific native speaker as a model in units of rhythm, phonemes, or words and calculate the differences. 
However, they require extensive speech data with detailed annotations or can only compare with one specific native speaker.
To overcome these problems, we propose a new language learning support system that calculates speech scores and detects mispronunciations by beginners based on a small amount of unannotated speech data without comparison to a specific person.
The proposed system uses deep learning--based speech processing to display the pronunciation score of the learner's speech and the difference/distance between the learner's and a group of models' pronunciation in an intuitively visual manner. 
Learners can gradually improve their pronunciation by eliminating differences and shortening the distance from the model until they become sufficiently proficient. 
Furthermore, since the pronunciation score and difference/distance are not calculated compared to specific sentences of a particular model, users are free to study the sentences they wish to study.
We also built an application to help non-native speakers learn English and confirmed that it can improve users' speech intelligibility.
\end{abstract}

\begin{IEEEkeywords}
  computer-aided language learning (CALL), computer-assisted pronunciation training (CAPT), audio and speech interfaces, deep learning, self-supervised learning 
\end{IEEEkeywords}

\section{Introduction}
Learning a non-native language can be challenging, especially in terms of speaking, because it differs from one's native language in many aspects, including vocabulary, grammar, phonetics, and accent. 
To learn a foreign language effectively, learners need to understand the pronunciation differences between themselves and such native speakers of the language they are learning.
The most common way for learners to understand the differences is in dictation, where a native speaker or expert evaluates how well the learner hears the speech~\cite{Dean1972,Derwing1997}.
However, this method requires considerable time and effort, limiting learners' learning opportunities.
To automate this method, this study proposes a language learning support system called DDSupport.
This system uses machine learning technology to calculate and visualize the user's pronunciation score and the difference/distance between the user's pronunciation and that of native speakers based on raw speech data from native and non-native speakers. 

Computer-aided pronunciation training (CAPT) systems, have been actively studied~\cite{Nancy2016}.
In previous studies, the differences between the learner's pronunciation and that of a native speaker have been calculated using indices such as intonation, rhythm, and phonemes that are presented to the user. 
Implementing such methods requires high-quality, manually produced, and annotated speech data. 
However, collecting such data is particularly difficult, representing a significant barrier to many language learning systems. 
Therefore, this study proposes a CAPT system that can be built using only unannotated utterance data. 
The system first classifies the utterances of native and non-native speakers using a deep learning--based classifier. 
It then calculates the score of the user's utterances by calibrating the output results of the classifier. 
Another issue is that to build such a deep learning--based discriminator, a large amount of speech data from non-native speakers of the target language as well as from native speakers is usually required. 
However, it is usually difficult to collect speech data from non-native speaker, whereas speech data from native speakers can be easily collected from the radio, TV, and video streaming services. 
Therefore, the proposed system uses recent self-supervised learning techniques~\cite{Shuo2022} to utilize native speakers' speech data efficiently that can be easily collected. 
In this way, our system can be built without using a large amount of speech data from non-native speakers.

\begin{figure}
  \includegraphics[width=0.5\textwidth]{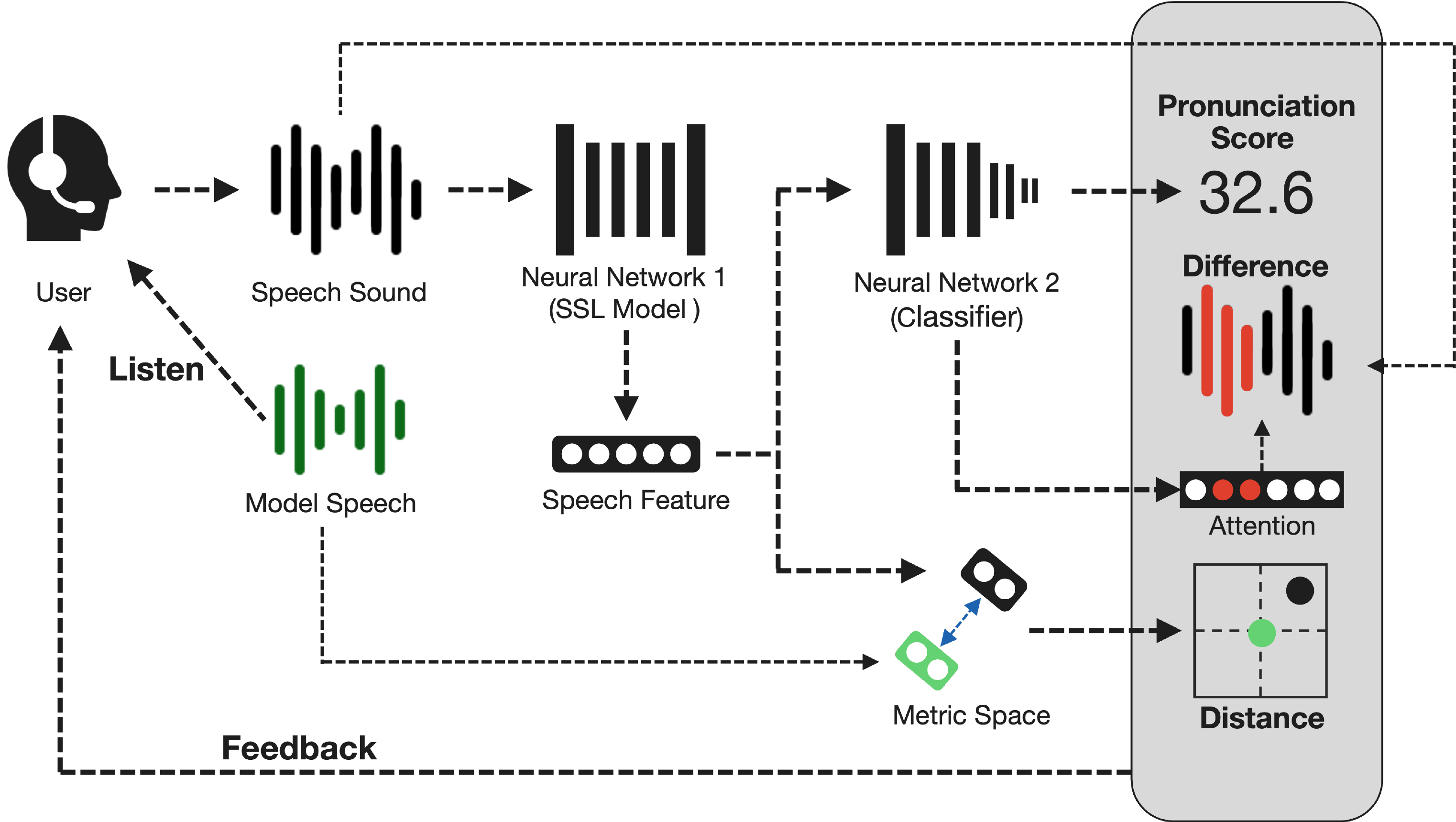}
  \caption{DDSupport architecture.}
  \label{fig:architecture}
\end{figure}

Moreover, to design a CAPT system with high learning effectiveness, it is essential not only to present the user's level of pronunciation in speech but also to provide easy-to-understand feedback on the difference between the user's speech and the model's speech. 
In existing CAPT systems, which detect errors in the user's speech based on such indicators as intonation, rhythm, and phonemes, it is sufficient to present these errors to the user. 
However, this strategy is not feasible for designing a CAPT system using unannotated data. 
Therefore, we provide two types of feedback that can be used even with unannotated data. 
First, areas where the user's utterance differs from the model's are highlighted in red on a waveform.  
Furthermore, to allow the user to see how far apart the user's utterance and the model's utterance are, they are represented as dots on the two-dimensional coordinates. 
One dot represents the user's utterance, and the other represents the model's utterance. 
Each is designed using attention, which allows the visualization of the deep learning model's decision criteria, and metric learning, which allows the design of a distance function from the data. 
We expect that the learner's utterance will approach the model by repeatedly modifying the utterance to reduce the distance between the two points, thus eliminating the difference in the waveform, until a satisfactory level of pronunciation is reached.
Furthermore, the three functions we propose (speech score, visualization of differences/distances) are calculated without comparison to a specific utterance of one particular native speaker, unlike in traditional CAPT systems. This means that users can practice speaking any sentences they like.

To evaluate the learning effectiveness of the proposed system, we implemented a support system for Japanese learners to learn to speak American English as a DDSupport application. 
To create this application, we used a large audio dataset of a native speaker of American English and a small unannotated dataset of a Japanese user speaking English. 
Subjects used the application to learn pronunciation, and their utterances were then rated by native speakers and through questionnaire-based self-evaluations.
The evaluation results showed that the proposed method effectively assisted in learning to speak a second language and improved the clarity of the learners' pronunciation. 
Moreover, the visualization of pronunciation score, difference, and distance was found to be an effective aid.

The main contributions of this paper are as follows:
\begin{itemize}
  \item A technical framework for assessing learners' speech proficiency and visualizing areas where they need improvement using unannotated speech data without comparison to specific native speakers. 
  \item A second Language Learning Support Application for Non-native English Speakers.
  \item A user study to evaluate the learning effectiveness of visually clear feedback.
\end{itemize}

\section{Related Work}
\subsection{Computer-Assisted Pronunciation Training}
CAPT systems are widely used to improve users' pronunciation when speaking a second language~\cite{Rogerson-Revell2021, Agarwal2019}. 
It is well known that corrective feedback enables efficient learning when learners attempt to imitate a given model utterance~\cite{Vries2011}. 
Thus, the main goal of CAPT is to record the learner's speech, detect and diagnose mispronunciations, and suggest ways to correct them. 
Recent research has mainly applied probabilistic models, such as machine learning models, to evaluate learners' speech proficiency and detect pronunciation error patterns. 
For example, some studies have used posterior probability scores based on hidden Markov models~\cite{Witt2000, Qian2010, Zhang2008} to detect pronunciation errors. 
Others have used decision trees~\cite{Li2016, Truong2004, Hongcui2009} or support-vector machines~\cite{Yoon2009, Yoon2010}. 
More recently, some studies have used deep neural networks~\cite{Kun2014, Hu2015, Hee2019} to improve detection performance. These systems effectively detect speaker errors but require detailed labeled speech data, such as phonemes, intonation, rhythm, and emphasis. 
In contrast, our method estimates proficiency and mispronunciation from native and non-native speaker speech data without detailed labeling.
Furthermore, the existing machine learning--based CAPT systems require large amounts of non-native speaker data as well as native speaker data. Our method effectively uses native speaker speech data, which are relatively easy to collect, and requires only small amounts of non-native speaker data.

\subsection{Interaction with CAPT}
Robertson et al. reported that user pronunciation improved regardless of pronunciation error detection performance~\cite{Robertson2018}.
They argue that CAPT research needs to be treated as human-computer interaction (HCI) problem, not just a speech information processing problem.
In other words, we need to consider not only how to evaluate user speech and calculate differences from model speech, but also how to present them and what kind of interaction to provide to the user.
For example, \cite{Kawahara2004} and \cite{Hongcui2009} presented users with errors related to prosodic features, such as syllable structure and stress, or vowel insertion or nonreduction.
Since our system uses data without human annotation, it is difficult to provide feedback on differences from model speakers in specific evaluation axes, such as phonemes, rhythm, and intonation. 
Therefore, we calculate the differences between the user and the exemplar directly at the speech level (not via specific evaluation axes) and present them on the waveform using changing colors. 
Some systems that use interactions, such as Tip Tap Tones~\cite{Darren2012}, SIAK~\cite{Reima2017}, and COP~\cite{Cristian2020}, allow users to learn a language through games. 
PTeacher~\cite{Yaohua2021} provides users with exaggerated audio and visual feedback, and Seiyuu-Seiyuu~\cite{Gabriel2017} facilitates the development of practical skills using a video learning interface. 
These systems have demonstrated the effectiveness of providing visual feedback and learning speaking through interaction. Inspired by these studies, our system was designed to facilitate language learning through a game-like experience by displaying two dots on a screen that are brought closer together as the user's speech improves.

\begin{figure*}[t]
  \begin{center}
  \includegraphics[width=1.0\textwidth]{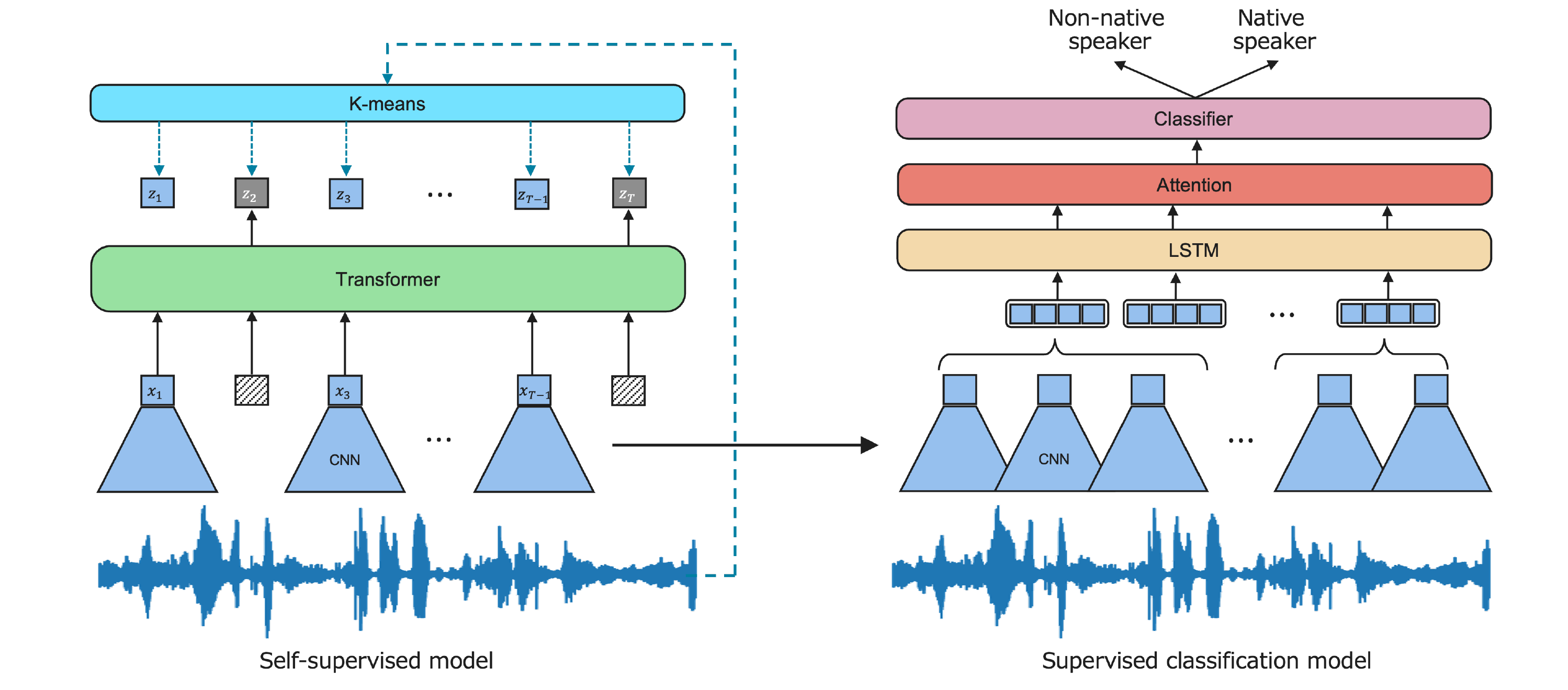}
  \caption{DDSupport pronunciation score determination.}
  \label{fig:proficiency}
  \end{center}
\end{figure*}

\section{Proposed System}
\subsection{Overview}
DDSupport has three main functions:
\begin{itemize}
  \item Pronunciation Score: a measure of how well the user speaks.
  \item Difference: different positions of the user's speech and a model's speech on the waveform.
  \item Distance: two points on the two-dimensional coordinates that indicate how far the user's speech is from the model's speech.
\end{itemize}
The proposed system uses mainly deep neural network techniques to achieve these functions, as shown in Fig.~\ref{fig:architecture}.
If the speech dataset contains values such as proficiency, difference, and distance as annotations, then a simple supervised classifier can be trained to build a model that finds these values in the user's speech.
However, such annotated data must usually be manually created, which is expensive. 
To address this problem, our system was designed to find the three desired values only from raw (unannotated) speech data from non-native and native speakers of the target language.
As shown in Fig.~\ref{fig:architecture}, the pronunciation score is calculated using the classification of the unannotated data, the difference is calculated using attention, and the distance is calculated using metric learning. 
A further problem is the difficulty of collecting speech data from non-native speakers.
In contrast, speech data from native speakers can be easily collected from various sources, such as the radio, TV, and video streaming services~\cite{Ardila2019,Vassil2015,Zen2019}. 
For this reason, the system was designed to effectively use large amounts of data from native speakers using self-supervised learning so that it does not need to collect large amounts of speech data from non-native speakers. 
\begin{description}
  \item[TP1] We need to train a deep learning model with minimal speech data from non-native speakers.
  \item[TP2] We need to obtain the desired values from speech data without human evaluation annotations.
\end{description}

\subsection{Pronunciation Score}
As shown in Fig.~\ref{fig:proficiency}, the model for calculating the pronunciation score of a user's speech is trained to classify speech data from non-native and native speakers of the target language.
We use LSTM~\cite{Hochreiter1997,Schuster1997} and the Softmax classifier to predict label $\hat{y}$ from a discrete set of classes $Y\in \{\rm native, non\mathchar`-native\}$ for speech audio $\mathcal{X}$.
The classifier uses the LSTM's hidden states $\bm{H} = [\bm{h}_{1}, \bm{h}_{2},  \ldots, \bm{h}_{T}]$ as input:
\begin{align}
  \begin{split}
    \hat{p}(y \mid \mathcal{X}) &= \operatorname{softmax}\left(\bm{W}^{(S)} (\bm{H}\bm{\alpha}^{\mathrm{T}}) + \bm{b}^{(S)}\right), \\
    \hat{y} &= \arg \max _{y} \hat{p}(y \mid \mathcal{X}),
  \end{split}
\end{align}
where $\bm{W}^{(S)}$ and $\bm{b}^{(S)}$ are trained parameter vectors,
and $\bm{\alpha}$ is the attention weight introduced in Section~\ref{sec:visualizing_speech_differences}.
Subsequently, the calibrated value~\cite{Nixon2019} that brings the output value of the model closer to the probability of belonging to each class is used as the proficiency value. 
Here, as shown in TP1, the problem of insufficient training data for the classes of non-native speakers arises.
Recently, self-supervised learning, a method in which the model provides teacher labels even if the data do not have them, has been actively studied~\cite{Schneider2019,Baevski2020,Shuo2022}.
Training such a model on a large amount of unlabeled data and fine-tuning it with a small amount of labeled data is known to effectively perform downstream tasks~\cite{Shu-wen2021}.
In our system, a self-supervised learning model is trained on a large amount of unlabeled speech data from native speakers, and then a binary classifier is trained on a small amount of labeled speech data from native and non-native speakers.
We use HuBERT~\cite{Hsu2021} as a self-supervised model for training with a massive amount of speech data from native speakers, obtaining latent representations $\mathcal{Z}$ from the input speech $\mathcal{X}$. 
Then, the time series data $[z^{\prime}_{1:k}, z^{\prime}_{k+1:2k}, \ldots, z^{\prime}_{T-k+1:T}]$ obtained by combining the representations $\mathcal{Z}$ at each step $k$ are fed into the LSTM. 

\subsection{Difference}\label{sec:visualizing_speech_differences}
\begin{figure}
  \includegraphics[width=0.5\textwidth]{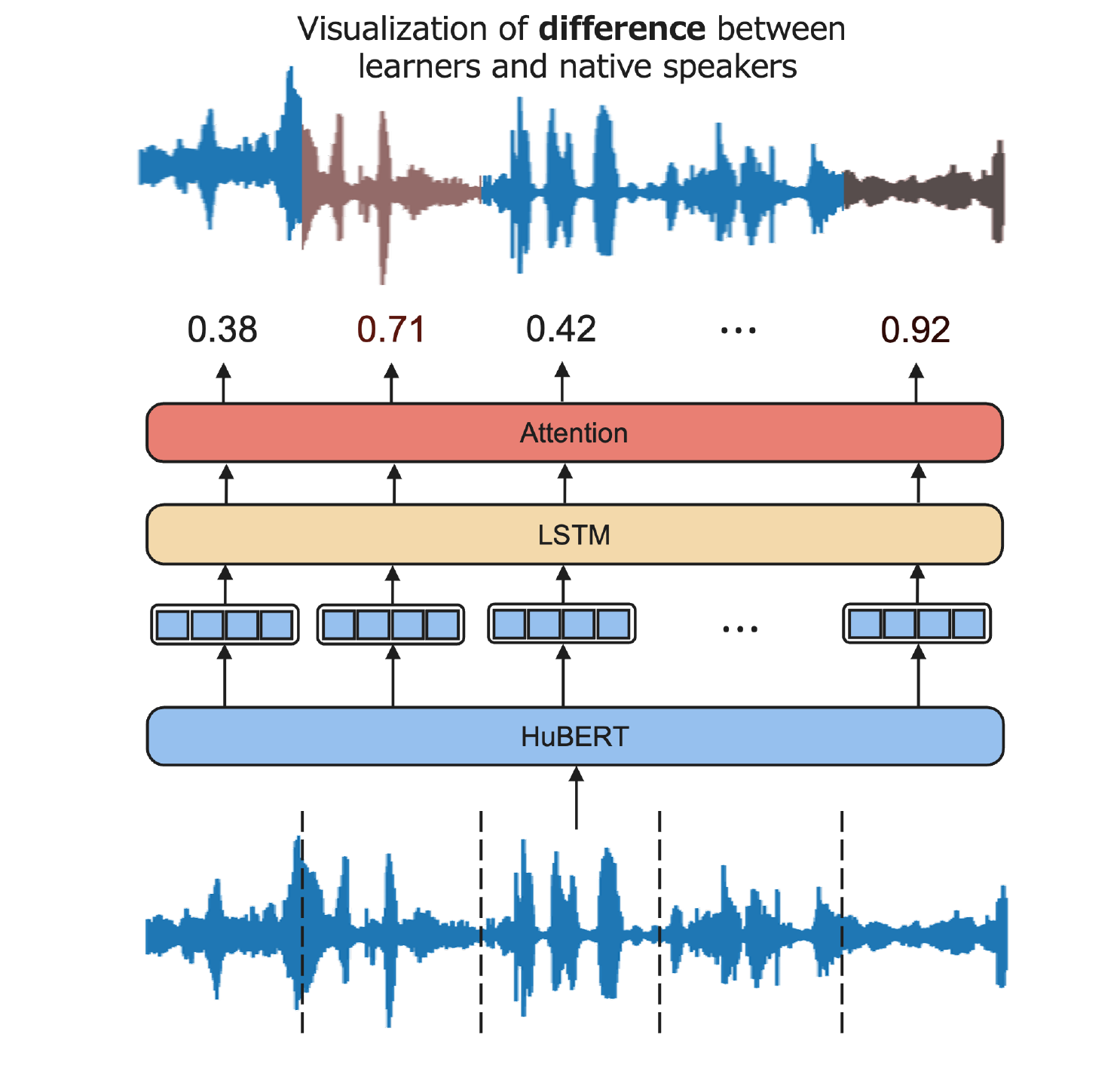}
  \caption{DDSupport difference visualization.}
  \label{fig:difference}
\end{figure}

Suppose that the non-native speakers' speech data have human-rated labels indicating which parts of the speech differ from that of native speakers. 
In this case, we can train a supervised learning model that inputs the speech data and outputs the parts that differ from the native speakers' speech. 
However, as stated in TP2, such labeled data are virtually unavailable. 
Therefore, our system uses the attention mechanism to calculate differences from unannotated speech data. 
Attention has been successfully used in various tasks, such as question answering, machine translation, and speech recognition~\cite{Bahdanau2014, Chorowski2015, Hermann2015},
to improve model performance and the interpretability of deep learning models.
Specifically, it visualizes where in the data the model focuses its attention when performing a particular task. 
As shown in Fig.~\ref{fig:difference}, we apply this mechanism to the classifier to determine the parts of the data on which the classifier focuses when it determines that the learner is not sufficiently proficient. 
The attention weight $\bm{\alpha}$, which indicates the importance of the speech each time step, can be obtained as follows:
\begin{align}
    \bm{M} &= \tanh(\bm{W}^{A1} \bm{H}), \;
    \bm{\alpha} = \operatorname{softmax}\left(\bm{W}^{A2} \bm{M}\right),
\end{align}
where $\bm{W}^{A1}$ and $\bm{W}^{A2}$ are trained parameter vectors.
The weight $\bm{\alpha}$ takes the value $[0,1]\in\mathbb{R}$.
This value is standardized, and the waveform in the corresponding area is displayed in red when it exceeds a certain threshold. 
The color is darker in segments that require greater attention attention to encourage the user to focus on them.

\subsection{Distance}
\begin{figure}
  \includegraphics[width=0.5\textwidth]{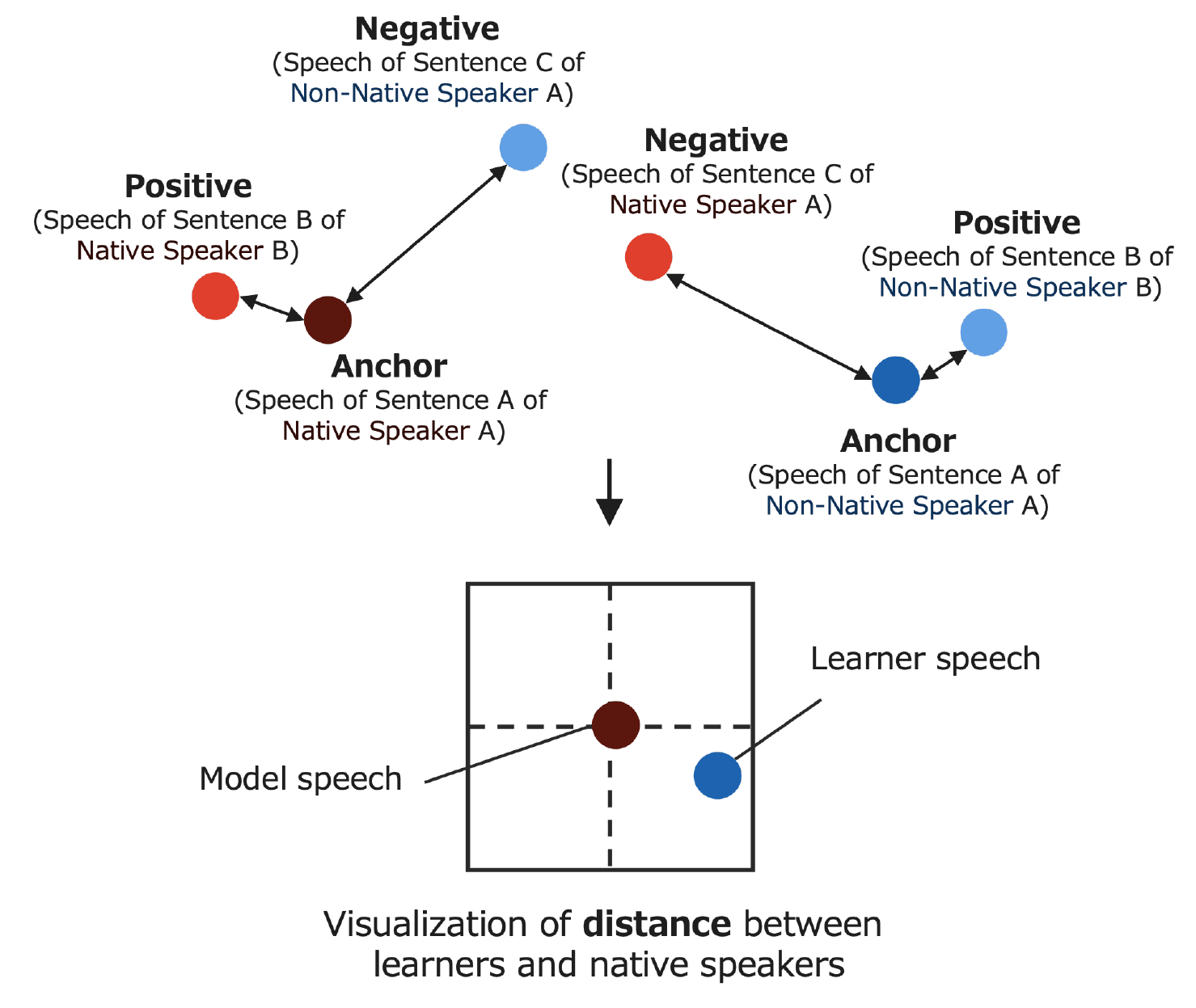}
  \caption{DDSupport distance visualization.}
  \label{fig:distance}
\end{figure}

Before calculating the speech distance between the user and the native speakers, it is necessary to design an optimal distance function to measure it.
One of the main approaches to learning the distance between data is triplet loss~\cite{Elad2015}.
Triplet loss is learned using the following three types of speech data, as shown in Fig.~\ref{fig:distance}:
\begin{itemize}
    \item Anchor: speech data from a non-native speaker.
    \item Positive: speech data from a non-native speaker different from anchor.
    \item Negative: speech data from a native speaker.
\end{itemize}
In obtaining the distance function, the distance from the anchor data to the positive data ($d_{a p}$) is trained to be smaller, while the distance from the anchor data to the negative data ($d_{a n}$) is trained to be larger:
\begin{align}
  L_{\text {triplet }}=\left[d_{a p}-d_{a n}+m\right]_{+},
\end{align}
where $m$ is the margin---that is, the desired difference between the anchor-positive and anchor-negative distances. 
The learned distance function is then used to calculate the two-dimensional vector difference between the user's speech data and the native speaker's speech data. 
In this case, the latter are the average of the native speaker's speech included in the data. 
Finally, the user's and native speaker's speech data are represented as points on the two-dimensional coordinates, with the native speaker's point at the center (Fig.~\ref{fig:distance}). 
In this way, the distance can be automatically calculated from the raw speech data, even if the data do not include a detailed human evaluation of how the native and non-native speech data differ, which is the challenge described in TP2.


\begin{figure*}[t]
  \centering
  \includegraphics[width=0.8\textwidth]{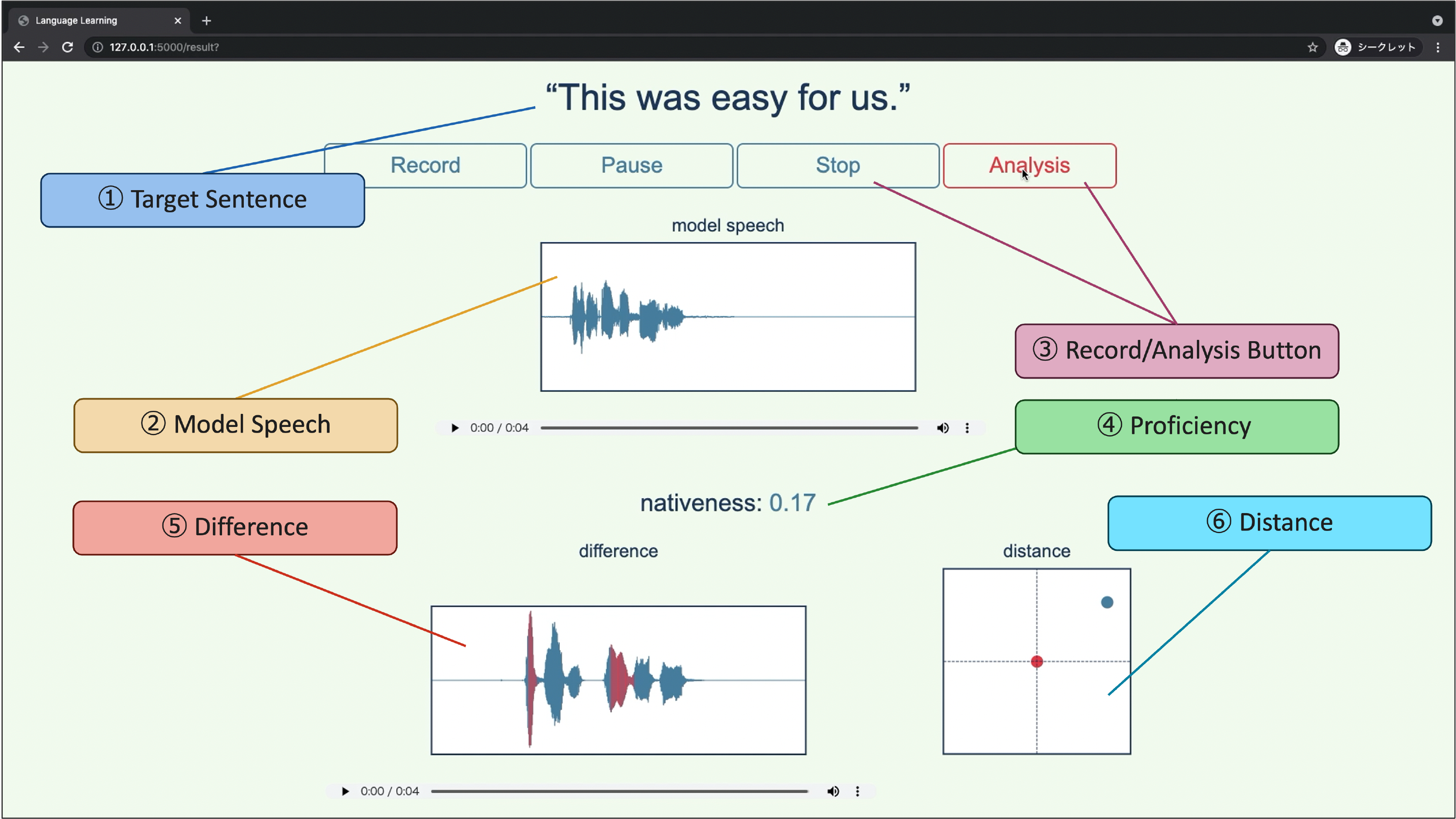}
  \caption{DDSupport user interface.}
  \label{fig:interface}
\end{figure*}

\section{English Learning Application}
\subsection{Implementation}
We developed a prototype DDSupport application to support Japanese people in learning to speak American English.
The model speakers are native speakers of standard American English as defined in \cite{Giegerich1992}.
Two datasets were used to create this prototype.
One was LibriSpeech (960~h)~\cite{Vassil2015}, which was used as a large unlabeled dataset of English speech to train HuBERT.
The other was UME-ERJ~\cite{Minematsu2001}, a database of English texts read by Japanese students and native speakers of standard American English, which was used as a small amount of labeled speech data.
UME-ERJ contains $806$ sentences spoken by $202$ Japanese and $20$ American speakers.
This dataset was used to train the classifier.
All audio files in the two datasets were standardized to a single-channel $16000$ $\mathrm{Hz}$ format and aligned to $4$ $\mathrm{s}$ in length.

We trained the model with HuBERT using the LibriSpeech data to obtain common speech features in English.
The attention layer used to visualize the differences was a $32$-dimensional hidden layer. 
We used LSTM with $128$-dimensional hidden layers and a linear layer with $512$ hidden layers for metric learning for distance visualization.
We used Dropout~\cite{Nitish2014} to train these models with a probability of $0.5$ to suppress overfitting.
Since UME-ERJ is unbalanced, containing more speech data from Japanese than from American English speakers, we used focal loss~\cite{Lin2017} as an error function to reduce the effect of the imbalance.
The optimization algorithm used Adam~\cite{Diederik2015}, with a batch size of $32$ and a learning rate of $10^{-4}$.
We also augmented the data to make them more tolerant of noise and to compensate for the scarcity of data. 
The data augmentation included adding Gaussian noise in the background, randomly increasing or decreasing the audio volume, randomly shifting the pitch, and silencing randomly selected audio parts.
These deep models were implemented using PyTorch~\cite{Paszke2019} as the backend and an NVIDIA GeForce 1080Ti GPU.
We then made it into a web application based on Flask, a microweb framework written in Python.

\subsection{Interface}
Fig.~\ref{fig:interface} shows the DDSupport interface. 
The system displays the target sentence that the user is about to learn to utter \ctext{1} and the model speech \ctext{2}.
By playing back \ctext{2}, the user can listen to the correct pronunciation of the target sentence. 
The user can input his/her speech into the system using the button \ctext{3}. 
The \textit{Pause} button can be used to interrupt the input. 
After the user has finished inputting his/her speech, the \textit{Analyze} button is used to analyze it.
When the \textit{Analyze} button is pressed, the degree nativeness of the user's speech \ctext{4} and the difference \ctext{5} and distance \ctext{6} between the model speech and the user's speech are displayed.
The difference is displayed in red on the waveform. 
The red area is the part that differs from the model speech, and the larger the difference, the darker the color.
The distance is displayed as two points on two-dimensional coordinates. 
The red point is the model's speech, and the blue point is the user's speech. 
The closer these points are, the closer the user's speech is to the model speech.

\begin{figure*}[t]
  \centering
  \includegraphics[width=0.9\textwidth]{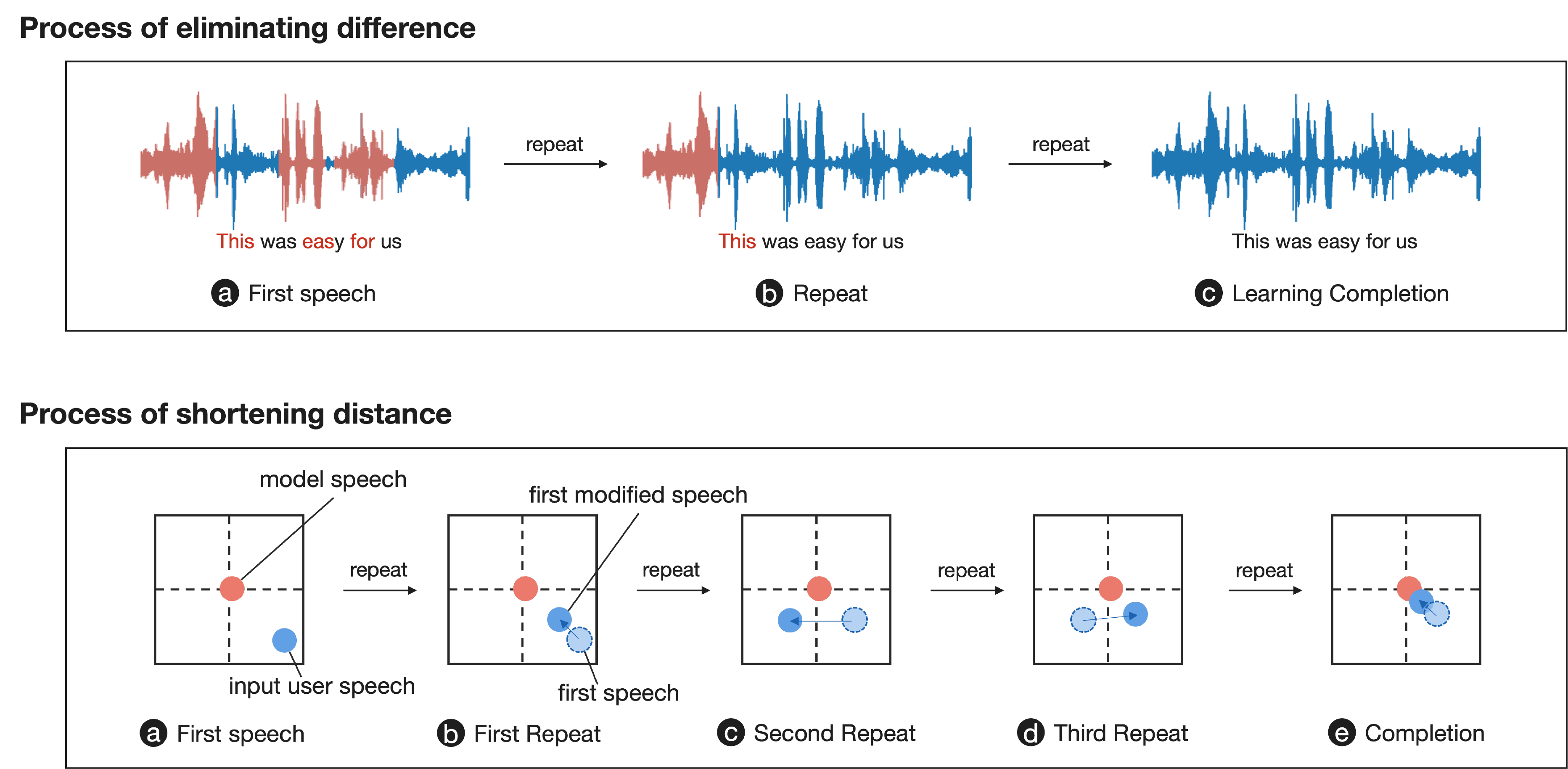}
  \caption{Examples of DDSupport interactions.}
  \label{fig:learning_process}
\end{figure*}

\subsection{Learning Process}
Here, we describe how a user can interact with the application successfully speak the target language.
The progress of the user's learning is shown in Fig.~\ref{fig:learning_process}.
The user speaks given sentences into the system.
Then, the system evaluates the user's utterance, gives a numerical value for the user's pronunciation score, and visually shows the difference and distance from the model utterance. 
The user repeatedly modifies his or her speech until the pronunciation score is satisfactory.
By repeating the sentence, the user can effectively modify his or her speech to eliminate the difference and shorten the distance. 
Each is described below.

\subsubsection{Eliminating the difference}
Fig.~\ref{fig:learning_process}~(upper panel) shows how the difference between the user's speech and the model's speech is eliminated as the user repeatedly modifies his or her speech.
In the waveform of the user's speech shown in the figure, the parts that differ from the model speech are indicated in red. 
When the user first speaks the sentence ``This was easy for us,'' the pronunciation of ``this,'' ``eas'' and ``for'' differes from that of the model speaker (\ctext{a}). 
When the user repeats the sentence to correct the red part shown by the system, the pronunciation of ``eas'' and ``for'' is improved (\ctext{b}). 
Finally, after correcting the pronunciation of ``this,'' which remains incorrect, the difference between the user's and the model's pronunciation disappears, as shown in \ctext{c}.

\subsubsection{Shortening the distance}
Fig.~\ref{fig:learning_process}~(lower panel) shows how the distance between the user's speech and the model's speech are brought closer together as the user repeatedly revises his or her speech.  
The model speech and the learner's speech are represented by dots on the two-dimensional coordinates in the diagram. 
The red dots represent the model speech, and the blue dots represent the learner's speech. 
The greater the distance between the dots, the more the user's speech differs from the model speech. 
The user changes the pronunciation so that the distance between these points becomes shorter.
If the user's dot moves closer to the model speech dot along a certain line every time the user repeats a sentence, the pronunciation change is correct (\ctext{a}$\rightarrow$\ctext{b} and \ctext{d}$\rightarrow$\ctext{e}). 
On the other hand, if the dot suddenly moves to an unexpected place when the user changes his or her pronunciation, this suggests that the user modified his or her speech incorrectly (\ctext{b}$\rightarrow$\ctext{c}).
Therefore, the user can modify their speech effectively to achieve a constant change (a change in which the user's dot approaches the model's dot in the same trajectory).

\section{Experiments}
We conducted a user study to evaluate the learning effectiveness of speaking practice using DDSupport.
We examined whether prototype application helped Japanese people improve their English pronunciation. 
Ten individuals aged 24--35 years participated in the experiment.
They were non-native speakers and had no long-term residence in an English-speaking country.

\subsection{Procedure}
This experiment compared DDSupport with elicited imitation (EI), a method of learning speaking by having the user listen to a model speech and imitate it several times. 
On the other hand, practice using our system provides feedback on the user's pronunciation score and difference and distance from model speech to help the user modify his or her speech accordingly. 
Each participant was presented with $10$ sentences from TIMIT~\cite{Garofolo1993}, a major English speech corpus, and learns to speak half of them using EI and the other half using DDSupport's method.
To ensure that the effect estimation would not be affected by the level of difficulty, the sentences were divided into five-sentence sets A and B, and half of the subjects learned A using DDSupport, and the other half learned B using DDSupport.

efore the experiment, the participants attended a training session to familiarize themselves with the two methods.
During the experiment, all participants' utterances were recorded, and each sentence's first and last utterances was used to estimate the learning effect.
Once the participants had completed the speaking practice with both methods, a brief questionnaire was administered regarding their user experiences. 

\subsection{Results}
\subsubsection{Learning Effect}
To compare the effectiveness of DDSupport with that of the standard EI method, we presented five American raters with two sets of pre- and post-practice sentence pairs and asked them to judge which sentence was more intelligible. 
The raters were blinded to the pre- and post-practice sentences.

\begin{figure}
  \includegraphics[width=0.5\textwidth]{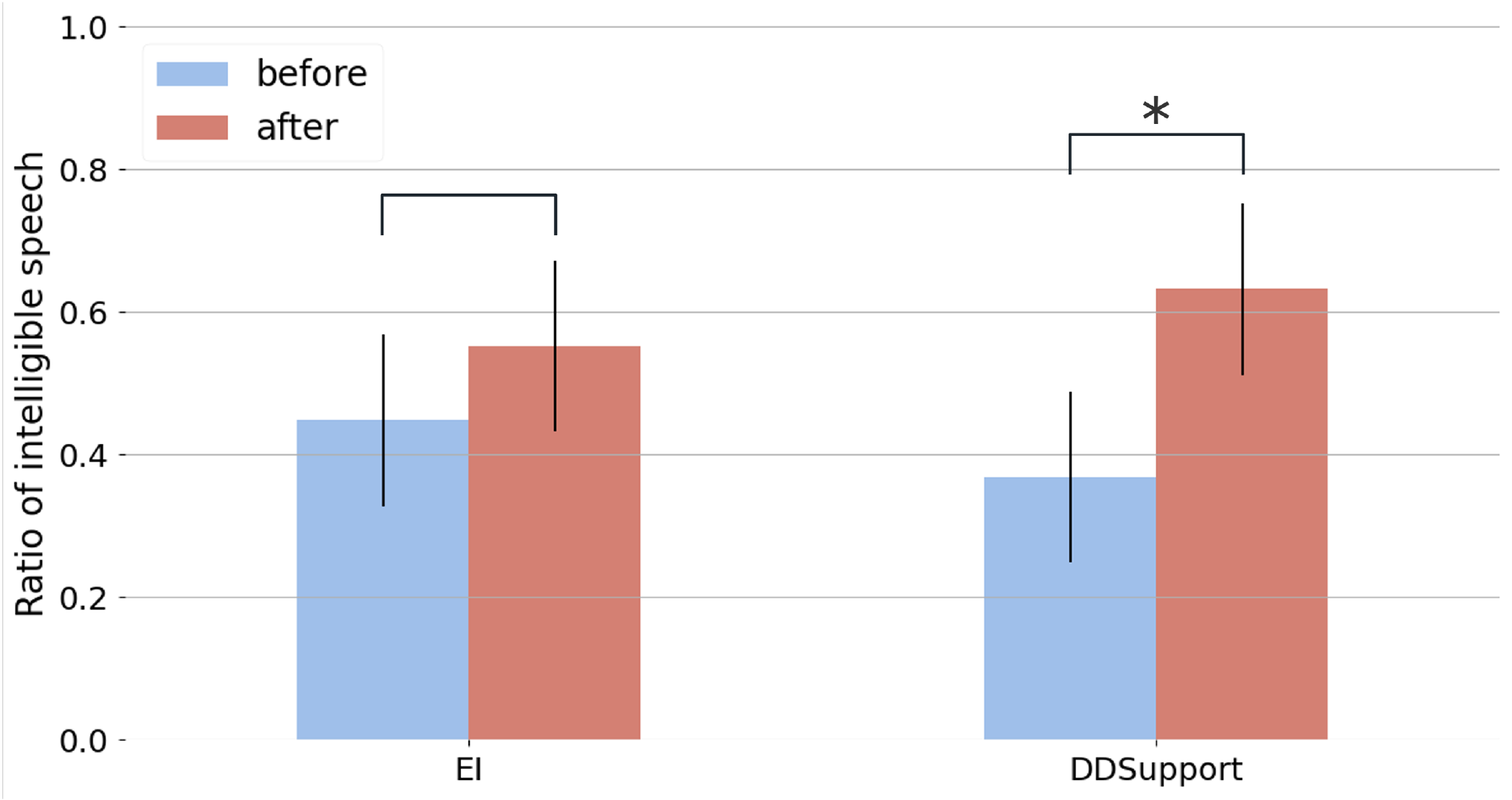}
  \caption{Comparison of post-practice intelligibility gains for users of DDSupport and standard EI method}
  \label{fig:user_study_1}
\end{figure}

The experimental results are shown in Fig.~\ref{fig:user_study_1}.
The y-axis indicates the percentage of the user's speech judged to be more intelligible when compared before and after learning.
The sentences practiced using both DDSupport and EI were found more intelligible that the pre-practice sentences.
However, the sentences practiced using DDSupport were found more intelligible than those practiced using EI.
Furthermore, unlike EI, DDSupport did not have overlapping pre- and post-practice confidence intervals, indicating a significant difference.

\subsubsection{User Experience and Questionnaires}
\begin{figure}
  \includegraphics[width=0.5\textwidth]{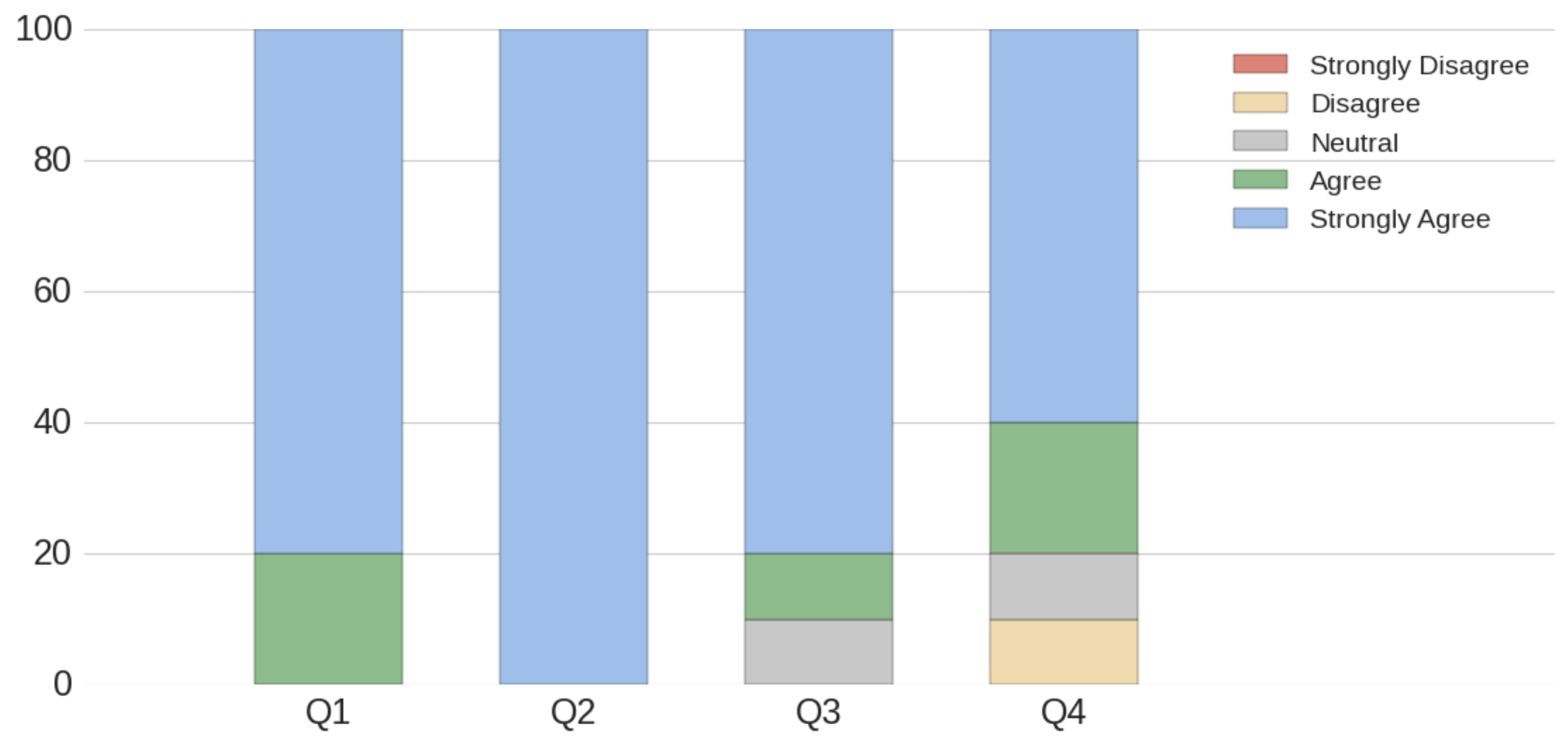}
  \caption{Questionnaire results}
  \label{fig:user_study_2}
\end{figure}

After the experiment, we conducted questionnaires with the following Likert scale questions about the DDSupport:
\begin{description}
  \item[Q1] I found it easy to use DDSupport.
  \item[Q2] I found the display of pronunciation score helpful.
  \item[Q3] I found the display of difference helpful.
  \item[Q4] I found the display of distance helpful.
\end{description}
For each question, we asked respondents to rate on a five-point scale of strongly agree, agree, neutral, disagree and strongly disagree.
The questionnaire results are shown in Fig.~\ref{fig:user_study_2}.
All participants found proposed system easy to use ("agree"/"strongly agree").
Moreover, all participants found the presentation of the pronunciation score helpful.
On the other hand, while the majority found the presentation of differences and distances useful, some answered "neutral" or "disagree" (one and two, respectively).
In summary, useful features in DDSupport's functional assessment are pronunciation scores, differences, and distance, in that order.

\section{Discussion}
The results of the learning effect of DDSupport, which was significantly stronger than that of the standard EI method of simply repeating utterances, and the results of the questionnaires confirm that this system is effective in helping users learn.
However, given some of the dissatisfaction with its functionality, it is clear that there are still improvements to be made in visualizing the distances and differences.
In particular, the quality of feedback to users may be improved by making the distance visualization axis more understandable, although it is expected to be very difficult to design a distance space using unlabeled data that is comprehensible to humans.
Also, for visualizing differences, other representations, such as text, can be co-labeled. 
Speech translation technology is maturing, and there is a growing number of easy-to-use packages~\cite{Wolf2019} that can be used for this purpose.
Furthermore, although we used publicly available datasets in this study, it is possible to design a personalized system using each user's speech data. 

We believe that this system can be used to support the acquisition of skills other than speaking.
It is possible to develop support systems for other skills based on speech information, such as musical instruments and singing.
In addition to audio representation, our system's self-supervised model could be replaced with a network that can handle other modalities, to support a wider range of skills.
For example, replacing it with the model for video representation used in \cite{Kaiming2020} and \cite{Ting2020}, which can handle video, could be used to visualize differences between professionals in sports and video production. 
Also, by replacing it with the model for text representation used in \cite{Yinhan2019} and \cite{Jacob2019}, which can handle text, it could be applied to support writing and programming skill acquisition.

\section{Conclusion}
In this paper, we propose a system that can support language learning by determining whether a user's speech is good and displaying the difference and distance between the user's speech and that of a native speaker. 
The system uses only unannotated speech data and does not require comparisons with specific native speakers. 
Thus, users can freely select sentences they wish to learn.
We also implemented the system as an application for learners whose native language is not English and confirmed that the learners' speech improves.
In the future, this method of evaluating a learner's skill level and providing guidance on where and how to improve is expected to be applied in various fields, such as sports and music.

\section*{Acknowledgement}
This work was supported by JST Moonshot R\&D Grant Number JPMJMS2012, JST CREST Grant Number JPMJCR17A3, and The University of Tokyo Human Augmentation Research Initiative.

\bibliographystyle{IEEEtran}
\bibliography{paper}

\end{document}